%% file: impetus.tex
\documentclass[acmsmall]{acmart}

\usepackage{algorithm}
\usepackage{algorithmic}
\usepackage{xac}

\AtBeginDocument{%
  \providecommand\BibTeX{{%
    \normalfont B\kern-0.5em{\scshape i\kern-0.25em b}\kern-0.8em\TeX}}}

\setcopyright{rightsretained}
\acmJournal{PACMHCI}
\acmYear{2021} \acmVolume{5} \acmNumber{CSCW1} \acmArticle{10} \acmMonth{4} \acmPrice{}\acmDOI{10.1145/3449084}

\begin{document}
\title{Lessons Learned from Designing an AI-Enabled Diagnosis Tool for Pathologists}

\author{Hongyan Gu}
\email{ghy@ucla.edu}
\affiliation{%
  \institution{University of California, Los Angeles, USA}
}

\author{Jingbin Huang}
\email{jihuang97@ucla.edu}
\affiliation{%
  \institution{University of California, Los Angeles, USA}
}

\author{Lauren Hung}
\email{hsienhuh@andrew.cmu.edu}
\affiliation{%
  \institution{Carnegie Mellon University, USA}
}

\author{Xiang `Anthony' Chen}
\email{xac@ucla.edu}
\affiliation{%
  \institution{University of California, Los Angeles, USA}
}

\renewcommand{\shortauthors}{Hongyan Gu, et al.}

\begin{abstract}
  \input{00_abstract.tex}
\end{abstract}



\maketitle

\input{lessons.tex}

\input{01_intro.tex}
\input{02_relatedwork.tex}
\input{03_walkthrough.tex}
\input{04_implementation.tex}
\input{05_study.tex}

\input{06_lessons.tex}
\input{07_discussion.tex}
\input{09_acknowledgement.tex}

\bibliographystyle{ACM-Reference-Format}
\bibliography{references,xai,xacpubs}

\received{June 2020}
\received[revised]{October 2020}
\received[accepted]{December 2020}

\appendix
\input{08_supplementary.tex}


\end{document}

%% file: 00_abstract.tex
Despite the promises of data-driven artificial intelligence (AI), little is known about how we can bridge the gulf between traditional physician-driven diagnosis and a plausible future of medicine automated by AI. Specifically, how can we involve AI usefully in physicians' diagnosis workflow given that most AI is still nascent and error-prone (\eg in digital pathology)? To explore this question, we first propose a series of collaborative techniques to engage human pathologists with AI given AI's capabilities and limitations, based on which we prototype Impetus --- a tool where an AI takes various degrees of initiatives to provide various forms of assistance to a pathologist in detecting tumors from histological slides. We summarize observations and lessons learned from a study with eight pathologists and discuss recommendations for future work on human-centered medical AI systems.

%% file: lessons.tex
\newcommand{\lone}[0] {
    \sdbox{
        To explain AI's guidance, suggestions and recommendations, the system should go beyond a one-size-fits-all concept and provide instance-specific details that allow a medical user to see evidence that leads to a recommendation.
    }{1}
}

\newcommand{\ltwo}[0] {
    \sdbox{
        Medical diagnosis is seldom a one-shot task, thus AI's recommendations need to continuously direct a medical user to filter and prioritize a large task space, taking into account new information extracted from a user's up-to-date input.
    }{2}
}

\newcommand{\lthree}[0] {
    \sdbox{
        Medical tasks are often time-critical, thus the benefits of AI's guidance, suggestions and recommendations need to be weighed by the amount of extra efforts incurred and the actionability of the provided information.
    }{3}
}

\newcommand{\lfour}[0] {
    \sdbox{
        To guide the examination process with prioritization, AI should help a medical user narrow in small regions of a large task space, as well as helping them filter out information within specific regions.
    }{4}
}

\newcommand{\lfive}[0] {
    \sdbox{
        It is possible for medical users to provide labels during their workflow with acceptable extra effort. However, the system should provide explicit feedback on how the model improves as a result, as a way to motivate and guide medical users' future inputs.
    }{5}
}

\newcommand{\lsix}[0] {
    \sdbox{
        Tasks treated equally by an AI might carry different weights to a medical user. Thus for medically high-staked tasks, AI should provide information to validate its confidence level.
    }{6}
}

%% file: 01_intro.tex
\section{Introduction}
\rev{The recent development of data-driven artificial intelligence (AI) is}{Recent advancements in machine learning techniques are} rejuvenating the use of \rev{AI}{artificial intelligence (AI)} in medicine that originally started over half a century ago. Enabled by data-driven statistical models, AI can already read X-Ray images \cite{Irvin2019,Cao2019} and analyze histological slides \cite{nalisnik2017interactive,campanella2019clinical} with a performance on par with human experts.

Despite their promises of automating diagnosis, existing medical AI models tend to be `imperfect' \cite{cai_human-centered_2019} --- there remain inherent limitations in the models' performance and generalizability. For example, in digital pathology, scanned tissue slides are processed by AI to detect tumor cells. The problem is that such histological data (\eg ovarian carcinoma) tends to have a high between-patient variance \cite{kobel2010diagnosis}; thus, a pre-trained model often struggles to generalize when deployed to a new set of patients.  At present, it remains underexplored how to integrate such `imperfect' AI usefully into physicians' existing workflow.

Researchers have long realized the limitation of using AI as a `Greek Oracle'. Miller and Masarie pointed out that a ``mixed-initiative system'' is mandatory whereby ``physician-user and the consultant program should interact symbiotically'' \cite{Miller1990}. Some research focused on mimicking how doctors think \cite{Shortliffe1993,Drew2013}, such as using an attention-guided approach to extract local regions of interest on a thoracic X-ray image \cite{guan2018diagnose}; others developed explainable models \cite{Caruana2015, Zhang2017} or system designs \cite{xie2020chexplain, Wang2019} that promote a doctor's awareness of AI's diagnosis process. Cai \etal developed a content-based image retrieval (CBIR) tool that allows a pathologist to search for similar cases by region, example, or concept \cite{cai_human-centered_2019}.
Yang \etal conducted \rev{field work}{fieldwork} to identify when and how AI can fit in the decision-making process of vascular assist device transplant \cite{Yang2016,Yang2019}. Despite such a growing body of work on mental models, explainability, CBIR tools, and field study, little has been done to answer the following question for medical AI: when AI is still nascent and error-prone, how can physicians still make use of such `imperfect' AI in their existing workflow of diagnosis?

To ground the exploration of this question, we focus on medical imaging --- the primary data sources in medicine \cite{Jiang2017}. Amongst various medical imaging techniques, histological data in digital pathology \cite{holzinger2017towards}, in particular, presents some of the most difficult challenges for achieving AI-automated diagnosis, thus serving as an ideal arena to explore the interactional relationship between physicians and AI.

Focusing on digital pathology as a point of departure, we propose a series of physician-AI collaboration techniques, based on which we prototype Impetus --- a tool where an AI aids a pathologist in histological slide tumor detection using multiple degrees of \rev{initiatives}{initiative}. 
Trained on a limited-sized dataset, our AI model cannot fully automate the examination process; instead, Impetus harnesses AI to
\one guide pathologists' attention to regions of major outliers, thus helping them prioritize the manual examination process;
\two use agile labeling to train and adapt itself on-the-fly by learning from pathologists; and
\three take initiatives appropriately for the level of performance confidence, from \rev{}{full} automation, to pre-filling diagnosis, and to defaulting back to manual examination.
We used the Impetus prototype as a medium to engage pathologists and observe how they perform diagnosis with AI involved in the process and elicit pathologists' qualitative reactions and feedback on the aforementioned collaborative techniques.
From work sessions with eight pathologists from a local medical center, we summarize lessons learned as follows.

\lone

\ltwo

\lthree

\lfour

\lfive

\lsix

Importantly, these lessons reveal what was unexpected as pathologists collaborated with AI using Impetus' techniques, which we further discuss as design recommendations for the future development of human-centered AI for medical imaging.

\subsection{Contributions}
Our contributions are as follows.
\begin{itemize} [leftmargin=0.25in]
    \item The first suite of interaction techniques in medical diagnosis that instantiate mixed-initiative principles \cite{Horvitz1999} for physicians to interact with AI with adaptive degree of initiatives based on AI's capabilities and limitations;
    \item A proof-of-concept system that embodies these techniques as an integrated \rev{diagnosis}{diagnostic} tool for pathologists to detect tumors from histological slides;
    \item A summary of observations and lessons learned from a study with eight pathologists that provides empirical evidence of employing mixed-initiative interaction in the medical imaging domain, thus informing future work on the design and development of human-centered AI systems.
\end{itemize}

%% file: 02_relatedwork.tex
\section{Related Work}
Our review of literature starts from a general body of cross-disciplinary work on human-AI interaction, gradually drills down to the (medical) imaging domain, and finally summarizes \rev{}{the} current status on digital pathology, which exemplifies the gap between traditional manual diagnosis and not-yet-available AI-enabled automation. 

\subsection{Human-AI Collaborative Work}
Since J. C. R. Licklider's vision of `man-machine symbiosis' \cite{licklider1960man}, bringing human and AI to work together collaboratively has been a long-standing challenge across multiple fields.

\rev{Recent work has been employing human-AI interaction to utilize human input to better and further reduce manual effort in repetitive routines.}{Recent work has been employing human-AI interaction to utilize human input better and reduce manual effort in repetitive routines.}
Specifically, Amershi \etal  propose a system that gives the user flexibility to provide better training examples in interactive concept learning \cite{amershi2009overview}. The system also grants users control over the training process: users could decide to stop training to avoid overfitting.
Chau \etal combine visualization, user interaction, and machine learning to guide users to explore correlations and understand the structure of large-scale network data \cite{chau2011apolo}.
Suh \etal show that classifier training with mixed-initiative teaching is advantageous over both computer-initiated and human-initiated counterparts. Specifically, mixed-initiative training could significantly reduce the labeling complexity across a broad spectrum of scenarios, from perfect, helpful teachers who always provide the most helpful teaching, to `naïve' teachers who give unhelpful labels \cite{suh2016label}.
Felix \etal propose a topic modeling system that could find unknown labels for a group of documents: by integrating human-driven label refinement and machine-driven label recommendations, the system enables analysts to explore, discover and formulate labels \cite{felix2018exploratory}.

Research \rev{also has}{has also} shown that human-AI collaboration can enhance domain-specific tasks.
For example, Forté enables designers to interactively express and refine sketch-based 2D design automated by topology optimization: specifically, a user can modify the optimization's result, which serves as input for the next iteration to reflect the user's intent \cite{chen2018forte}.
Nguyen \etal combine human knowledge, automated information retrieval, and machine learning in a mixed-initiative approach to conduct fact-checking \cite{nguyen2018believe}. However, in the medical domain, Yang \etal point out that the involvement of the machine-initiative often fails to assist physicians in clinical practices \cite{Yang2016}. In particular, ``the disruptive, time-consuming'' machine work would ``conflict with the chaotic nature of clinical work'', which undermines the mental need \rev{of}{for} machine support for medical users. Building on this finding, Yang \etal further adapted the concept of unremarkable computing, where doctors could seek \rev{for}{} machines' AI support on-demand to reduce AI's disruptive behavior\cite{Yang2019}.

To democratize the design of human-AI collaboration, Horvitz articulated a series of principles of mixed-initiative interaction via \rev{an}{the} example of an email-to-calendar scheduling agent \cite{Horvitz1999}. Insights from Horvitz's work were renewed in a recent paper by Amershi \etal, which proposes and validates 18 guidelines for human-AI interaction, which includes, for instance, ``support efficient correction'', ``make clear why the system did and what it did'', ``convey the consequences of user actions'' \cite{amershi2019guidelines}.
However, it is ``uniquely difficult'' \cite{yang2020re} to implement those guidelines. Due to the ``uncertainty surrounding AI's capabilities'' and ``complexity in AI's output'', it is hard for humans to control and trust AI systems \cite{yang2020re}. Given such challenges in system design, human-AI collaboration guidelines in the medical domain focus on building human-centered systems that revolve around AI \textit{as-is}. When AI makes mistakes, instead of modifying or correcting the model, current work emphasizes \rev{on}{} informing users of AI's pitfalls to ensure transparency \cite{cai2019hello}. To the best of our knowledge, there remains a lack of research on discussing how human-AI collaboration could provide new knowledge to AI in the medical domain.

Going beyond collaborating with AI as-is, our work is grounded in a proof-of-concept system situated in a specific application context, through which we found that the sheer amount of high-resolution medical imaging data posts unique challenges: \one how to filter AI's analyses on such high-resolution data and communicate actionable information to advance a differential diagnosis that is by nature iterative and \two how to cost-effectively incorporate pathologists' input as new knowledge to AI without taxing them with labeling a large amount of data.

\subsection{Data-Driven Digital Image Processing}
Imaging provides an abundant source of clinical data in medicine \cite{Jiang2017}. Furthermore, data-driven AI has served as a powerful toolkit for processing digital images (\eg the recent breakthrough in deep models for examining chest X-ray \cite{Irvin2019}). Remarkably, the increase of \rev{data collections}{data availability} in digital pathology \cite{weinstein2013cancer, veta2019predicting, roux2013mitosis, litjens20181399} has triggered a recent surge of data-driven techniques in a board range of applications, such as carcinoma detection \cite{araujo2017classification, bejnordi2017diagnostic, bardou2018classification}, histological features detection \cite{cirecsan2013mitosis, veta2015assessment}, and tumor grading \cite{ertosun2015automated, arvaniti2018automated}.

While some works mentioned above have reported expert-level performance of AI models, human involvement remains indispensable, primarily in the provision of training labels.
However, labeling medical data is a non-trivial task, since it usually suffers from a high variance in tissue appearance \cite{kobel2010diagnosis}, subjectivity in medical guidelines \cite{cai2019hello}, and, most importantly, a rare availability of trained specialists \cite{schaekermann2020expert}. Those barriers result in a low throughput in the medical image processing pipeline.
Schaekermann \etal try to break this dilemma by training more-available general clinicians --- their comprehension of difficult cases could be improved by being exposed to specialist adjudication feedback \cite{schaekermann2020expert}. However, the validation study shows that labeling performance does not increase significantly, since training humans without comprehensive, informative teaching would incur confusion.

In lieu of training humans, another way to reduce labeling cognitive workload is by using human labels more efficiently. Specifically, multiple works \cite{sommer2011ilastik, nalisnik2017interactive, zhu2014scalable} have employed the concept of active learning \cite{settles2009active}, where machines could learn from previous input iteratively and recommend the most uncertain cases to users for annotation. For example, Sommer \etal propose Ilastik for cell segmentation model training, which allows users to annotate by drawing strokes over cells \cite{sommer2011ilastik}. However, the microscopic nature of \rev{such labels}{stroke annotations} demands much user effort to achieve \rev{gigapixel whole-slide level performance}{good performance on the gigapixel whole-slide level}. Nalisnik \etal implement HistomicsML to perform nucleus identification with a random forest classifier, which dynamically recommends the most uncertain patches for annotation in each training iteration \cite{nalisnik2017interactive}. Going beyond selecting the most uncertain samples, Zhu \etal add a \rev{diverse}{diversity} constraint to reduce recommended samples' over-concentration in a localized area \cite{zhu2014scalable}. However, due to the high variation of histological features, merely relying on limited human annotations without exposing the model to a comprehensive training set would cause bias. This would result in a dispersion of sensitive false-positives across the whole-slide image, and humans would lose trust in machines while being overwhelmed with false-positive information.

To address this `low information input' issue without significantly increasing human burden, recent works \cite{xu2012multiple, ilse2018attention, campanella2019clinical} have applied multiple instance learning (MIL) \cite{zhou2004multi, babenko2008multiple} approaches to digital pathology, where MIL learners could even learn from whole slide-level labels, thus dispensing with the need for users to label at the pixel-level. For example, Xu \etal employ MIL \cite{xu2012multiple} to train a patch-level classifier to identify colon cancer based on slide-level annotations. Ilse \etal  use an attention-based deep MIL approach to learn a Convolutional Neural Network (CNN) with image-level annotations. The learned CNN can highlight breast or colon cancer areas, with reported performance on par with other MIL approaches without sacrificing interpretability \cite{ilse2018attention}. Campanella \etal combine MIL with Recurrent Neural Network (RNN) techniques and train with slide-level diagnosis for prostate cancer, basal cell carcinoma, and breast cancer metastases \cite{campanella2019clinical}. However, it usually requires a large amount of slide-level annotations for training (for example, \cite{campanella2019clinical} \rev{used}{collected a database of } >44,000 slides from >15,000 patients \rev{for training}{}); otherwise, there is often a performance drop compared to using strongly-supervised labels on the same set of slides.

Different from previous work, we focus on training machines with reduced human burden, addressing the issue of adopting a generic, pre-trained model on a new set of data. Specifically, we apply a \rev{pre-trained CNN that has been previously trained on a larger dataset as a feature extractor first}{ CNN feature extractor that has been previously trained on a larger dataset}. The pre-trained CNN would generate a local embedding for the new data, which addresses the `low information input' in traditional active learning techniques. Then, instead of training an MIL learner end-to-end, we implement an MIL learner that can learn jointly from the data embedding and user input labels. This training scheme harnesses the benefit of reducing human annotation labels with MIL technique, while avoiding having to retrain an MIL learner from scratch.

\subsection{Interactive Tools for Digital Pathology}

Digital pathology, similar to biology research, often deals with high-resolution, visually challenging images. 
Beyond involving data-driven models trained by domain experts, tools that allow pathologists to define, explore, and decide upon clinical or research problems are also needed.

ImageJ \cite{schneider2012nih} and its distributions \cite{schindelin2012fiji, rueden2017imagej2, collins2007imagej} are the primary scientific image analysis tools for digital pathology. They not only allow pathology researchers to \rev{perfrom}{perform} image operations (\eg cropping, measuring, editing), but support a wide range of plugins for immunohistochemistry (IHC) and fluorescence analysis \cite{jensen2013quantitative}. Besides ImageJ, multiple interactive tools have been proposed to aid medical users to automate whole-slide image (WSI) analysis without coding, covering domains of phenotype analysis \cite{carpenter2006cellprofiler}, segmentation \cite{saltz2017containerized}, and IHC screening \cite{martel2017image}. Recent research \cite{Yang2016, Yang2019, DBLP:conf/iui/XieCG19, xie2020chexplain} suggests that, besides reasoning with medical data, the design of a diagnosis tool often needs to \rev{take into consideration}{consider} physicians' established workflow and other domain-specific behaviors. Some digital pathology tools further consider the social background of clinical use and support database constructions \cite{gutman2017digital} and remote collaborations \cite{maree2016collaborative}.

When it comes to diagnosis, Blois considers the relationship between physicians and computer systems as a `funnel' \rev{}{as in} \fgref{funnel} \cite{doi:10.1056/NEJM198007243030405}: from A to B, a physician needs to gather information, formulate hypotheses, perform tests and gradually arrive at a diagnosis. Many of the aforementioned tools tend to focus on and perform well at or near Point B, \eg performing a physician-defined measuring task. In comparison, we design Impetus \rev{also}{} to support physicians' examination at or near Point A \rev{}{as well}, \eg suggesting where on a WSI to examine first.

Apart from this work, some recent interactive tools have also targeted supporting physicians in a border context. Specifically, Hegde \etal propose a deep-learning reverse image searching tool to help pathologists search for images with similar tissue type, histological feature, or disease state \cite{hegde2019similar}. Given a specific image patch, the tool first calculates image embeddings through a deep-learning model. Then, it returns the images' nearest neighbors in \rev{}{the} embedding space as a searching result. Building upon the searching algorithm, Cai \etal build a CBIR system that allows users to retrieve similar image patches from a database. Meanwhile, users could refine the searching results on-the-fly by regions, examples, or pre-defined histological concepts to deal with the searching algorithm's imperfection behavior \cite{cai_human-centered_2019}. Our work differs from \cite{hegde2019similar, cai_human-centered_2019} in two ways: \one instead of enabling users to \textit{adjust} imperfect \textit{searching} results while leaving the underlying model untouched, our tool seeks to address imperfect \textit{carcinoma detection} algorithm by cost-effectively incorporating pathologists' input; \two due to the difference in tasks (retrieval \textit{vs}. diagnosis), our work naturally goes beyond concentrating a localized region and discovers suspicious patterns with spatial structures at a whole-slide level, giving users a global view of the model behavior.

%% file: 03_walkthrough.tex
\section{Impetus: an AI-Enabled Tool for Pathologists}
Before we unfold our design process in the next section, we first \rev{introduce the background of}{give a background introduction on} digital pathology and the motivation to involve AI. We then walk through Impetus's scenario to present a high-level overview of how the tool works with pathologists.

\subsection{Background on Digital Pathology}
Central to digital pathology are whole-slide images or WSI for short. WSIs are produced by high-speed slide scanners that digitalize the glass slide at very high resolution, resulting in gigapixel images \cite{litjens20181399}.
Due to its large and high visual variance, a WSI cannot be directly fed into a model (such as a CNN) for classification. A WSI is usually divided into small patches, which are then classified by a CNN model (\rev{Figre}{Figure} \ref{fig:flow_chart}(a)). These patch level predictions can then be assembled to create a tumor probability heatmap, from which a pathologist can derive a whole-slide level diagnosis. 

In our study, we used a dataset containing Hematoxylin and Eosin (H\&E) stained sentinel lymph node (SLN) sections of breast cancer patients \cite{litjens20181399}. The diagnosis of such specimens contains four main categories \cite{apple2016sentinel}:

\begin{itemize}
    \item {\bf Isolated tumor cells (ITC)} if the node contains \rev{}{a} single tumor cell or cell deposits that are no larger than 0.2 mm or contain fewer than 200 cells;
    \item {\bf Micro} if containing metastasis greater than 0.2 mm or more than 200 cells;
    \item {\bf Macro} if containing metastasis greater than 2 mm;
    \item {\bf Negative} if containing no tumor cells.
\end{itemize}

\subsection{Promises \& Challenges of AI for Digital Pathology}
Digital pathology transforms traditional microscopic analysis of histological slides into high-resolution digital visualization \cite{holzinger2017towards}. Digital pathology allows pathologists to investigate fine-grained pathological information, transfer previously-learned knowledge to new tasks \cite{holzinger2017towards}, and, most importantly, \rev{to}{} leverage the recent development of data-driven AI to augment their visual analytical tasks.

However, the main challenge for digital pathology is that, unlike other imaging modalities (\eg X-Ray, CT), histological data (\eg ovarian carcinoma) tends to have a high between-patient variance \cite{kobel2010diagnosis}; thus, a pre-trained model often struggles to generalize when deployed to a new set of patients. Such an uncertainty of performance creates a barrier that prevents AI from being adopted to assist diagnosis in digital pathology.

To overcome this barrier, one solution is to improve the machine learning model by training it on a sufficiently large amount of patient data using cost-effective labeling and learning schemes \cite{nalisnik2017interactive, campanella2019clinical}. However, such a `big data' approach attempts to close the gap by (marginally) improving AI's performance, while ignoring the opportunity to engage human physicians. As a result, efforts are often bound to repeat the `Greek Oracle' pitfall pointed out by Miller and Masarie almost three decades ago \cite{Miller1990}. The focus of our paper is to explore and study the oft-missed opportunity of combining physicians with an `imperfect' AI: rather than awaiting AI to be fully automatable one day, how can we make use of its capability with limitations today?

Below we describe a scenario walkthrough of Impetus --- a tool that explores how AI --- without yet the ability to diagnose fully-automated --- can still empower pathologists by becoming an integral part of their workflow.

\subsection{Scenario Walkthrough}
The user of Impetus, a pathologist, starts \rev{the diagnosis of}{diagnosing} a patent's case by importing multiple Whole Slide Images (WSI) of the patient into Impetus. 

\fgw{walkthrough}{walkthrough}{0.38}{Key interactive features of Impetus: (a) as a pathologist loads a whole slide image, AI highlights areas of interest identified by outlier detection, shown as two yellow recommended boxes. (b) Agile labeling: a pathologist can drag and click to provide a label that can be \rev{used}{employed} to train the AI's model. (c) Diagnosis dialogue, pre-filled with AI's diagnosis, allows the pathologist to either confirm or disregard and proceed with manual diagnosis.}

First, the pathologist's attention is drawn to the two boxes generated by the AI, which encompass regions of patches that visually appear to be `outliers' from the majority of cells (\fgref{walkthrough}(a)), which suggests that these patches are likely to be tumor-positive. 
With these automatic recommendations, Impetus alleviates the pathologist's burden of navigating a large, high-resolution image and having to go through a large number of areas that might or might not be as tumor-characteristic as the recommended regions. 

Next, the pathologist performs diagnosis by marking each recommended region as either `tumor' or `normal', and continues to marquee-select and label a few more regions on the WSI (\fgref{walkthrough}(b)). As the pathologist makes such \rev{diagnoses}{selections}, their input is also collected by the back end AI and used as labels to adapt the model better to align itself with the pathologist's domain knowledge.

Based on these diagnostic inputs and revisions from the pathologist, Impetus immediately adapts the underlying AI model accordingly. 
In contrast to conventional data labeling tasks, Impetus' agile labeling is designed to be lightweight and \rev{able to}{can} learn from pathologists' input of coarsely marked regions without having to trace a precise contour of a tumor region. In this way, Impetus allows pathologists to agilely train an AI model as a natural and integral part of their existing workflow without incurring extra effort.

As the pathologist \rev{diagnoses}{annotates} more WSIs (which also trains the AI), they notice that some new slides are already marked as `diagnosed' --- AI takes the initiative to diagnose slides that it feels highly confident about. Thus the pathologist skips ahead to see other unlabeled slides, some of which, have pre-filled diagnosis dialogues (\fgref{walkthrough}(c)). In such cases, the pathologist examines the WSI to verify the AI's hypothesis. In the rest of the WSIs, the AI almost becomes invisible (due to a lack of confidence), and the pathologist proceeds to to finish the diagnostic tasks manually.

The above scenario demonstrates how an `imperfect' AI can still benefit a pathologist without necessarily automating the user's existing workflow: recommendation boxes suggestively prioritize pathologists' manual searching process (\fgref{walkthrough}(a)), agile labeling adapts AI while minimizing the extra effort from the pathologists (\fgref{walkthrough}(b)), and as AI attempts to improve itself, it handles cases with different degrees of initiatives --- from \rev{fully}{full} automation to pre-filling plausible results to remaining complete `invisible'---based on its confidence (\fgref{walkthrough}(c)).

%% file: 04_implementation.tex
\section{Design and Implementation}

Below we first describe the design process then detail the specific interaction techniques and their implementation in Impetus.

\subsection{Overview of the Design Process: Empirical \& Theoretical Grounding}
The design of Impetus is grounded in both empirical evidence and principles drawn from literature.

On the empirical side, we co-designed Impetus with our pathologist collaborator. Specifically, we learned that one major challenge for pathologists is efficiently and effectively navigating large, high-resolution WSIs. This suggests that AI, besides making diagnosis, can usefully serve to guide pathologists to navigate complex and high-resolution image space. We detail this design in {\S}4.2.

On the theoretical side, Impetus goes beyond the singular objective of automation by offering a spectrum of AI-enabled assistance. As pointed out by Blois' seminal paper \cite{doi:10.1056/NEJM198007243030405}, a physician's differential diagnosis process is similar to a funnel, starting with a broad exploration of plausible conditions and gradually rule out less likely possibilities as more evidence (\eg test results) is gathered until finally a single most probable conclusion can be drawn. According to Blois, AI has been canonically developed to optimize Point B, where a computer program can deterministically confirm whether a patient has a certain disease given all the evidence. As Blois foresaw, a recent development of AI starts to exhibit capabilities towards Point A, \eg Stanford's CheXpert produces likelihoods of 10+ thoracic diseases based on a chest X-ray image \cite{Irvin2019}. Similarly, Impetus also aims at ``reaching Point A'' by enabling pathologists' initial exploration with recommended regions.

\fg{funnel}{funnel_alt2}{0.6}{A physician's differential diagnosis process is similar to a funnel, starting with a broad exploration of plausible conditions and gradually rule out less likely possibilities as more evidence (\eg test results) is gathered until finally a single most probable conclusion can be drawn. Beyond mixed-initiatively automating certain diagnosis (near Point B), Impetus also supports exploration near Point A by enabling pathologists' initial exploration with recommended regions. Image modified based on Blois \cite{doi:10.1056/NEJM198007243030405}.}

Overall, Impetus provides the first suite of interaction techniques in the medical imaging domain  that instantiates mixed-initiative principles \cite{Horvitz1999} for physicians to interact with AI with \rev{}{an} adaptive degree of initiatives based on AI's capabilities and limitations. Specifically, we focus on the following principles in \cite{Horvitz1999}:

\begin{itemize} [leftmargin=0.25in]
    \item {\it Scoping precision of service to match uncertainty.}
    We first design a rule-based algorithm to identify three levels of uncertainty in AI's performance given a WSI (detailed in {\S4.4}), based on which we then design the corresponding AI-initiated action appropriate for each level of uncertainty (Table~\ref{tb:ai_initiatives}).
    \item {\it Providing mechanisms for efficient agent-user collaboration to refine results.}
    For each AI-initiated action, we also design mechanisms to introduce physician-initiated actions aimed at confirming, refining, or even overriding AI's results (Table~\ref{tb:ai_initiatives}).
    Further, we extend this principle by leveraging physician-initiated input for `machine teaching' \cite{simard2017machine}, \ie an agile labeling technique to dynamically adapt an AI by retraining it with examples of how a physician interpret a patient's histological data (detailed in {\S}4.3).
\end{itemize}

\begin{table*}[t]
    \small
\centering
\caption{Spectrum of human and AI initiatives at different AI confidence levels.}
\begin{tabular}[t]{lp{150pt}p{150pt}}
\toprule
\textbf{AI Confidence}&\textbf{AI-Initiated Action}&\textbf{Physician-Initiated Action}\\
\midrule
High&
Performing diagnosis automatically in the background; marking WSIs as diagnosed &
Doing nothing and accepts AI's results; can re-open a WSI to overwrite AI's result\\

$\uparrow$ &
Pre-filling the diagnosis box without directly labeling the WSI&
Performing diagnosis with help from AI predictions; confirming or correcting the pre-filled dialogue \\

Low&
Showing original WSI by default to prompt for manual diagnosis&
Performing diagnosis with little input from AI\\
\bottomrule
\end{tabular}
\label{tb:ai_initiatives}
\end{table*}%

\subsection{AI Guiding Pathologists' Attention to Regions of Major Outliers}
\label{subc_outlier}
\begin{figure}
    \centering
    \includegraphics[width=0.5\linewidth]{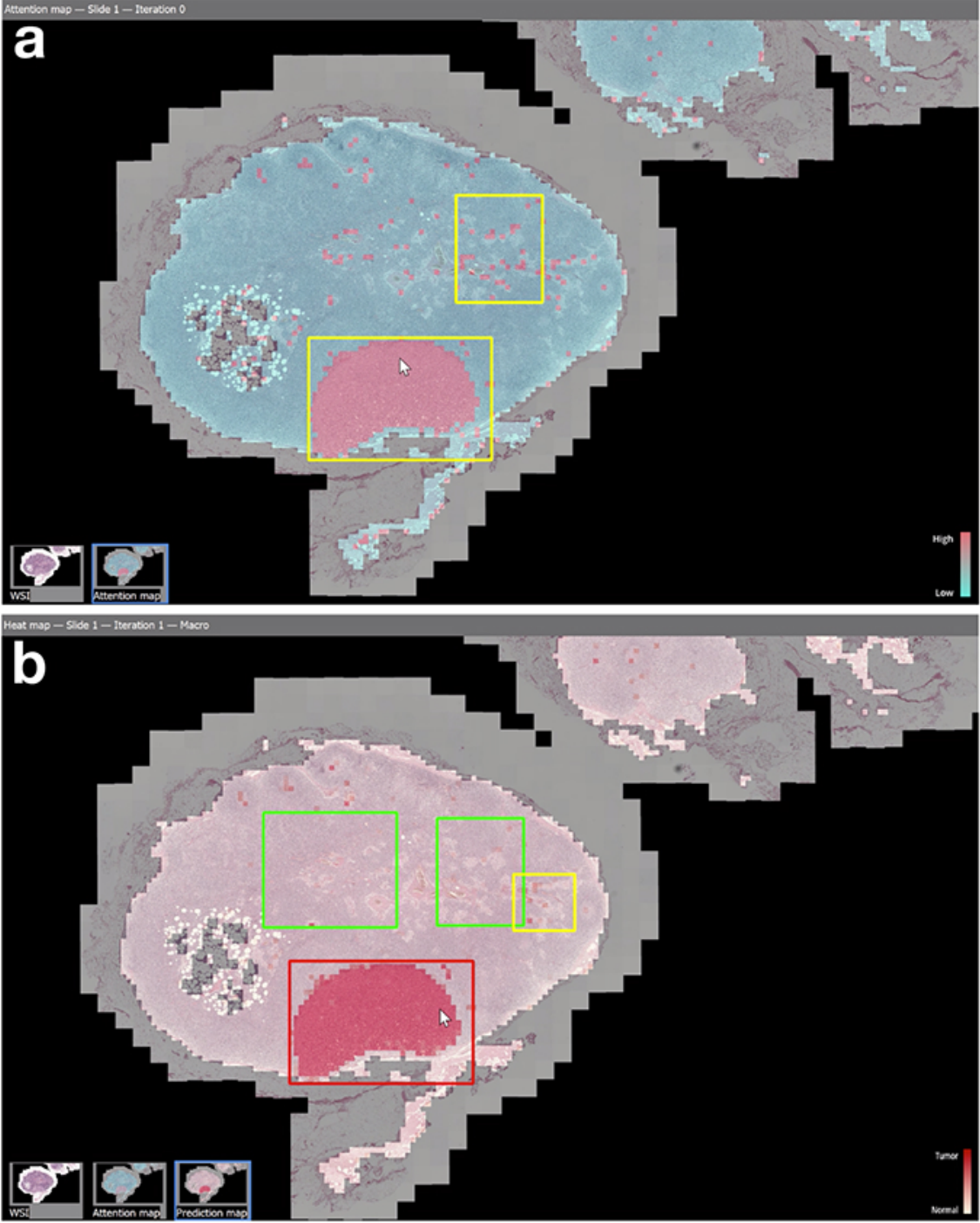}
    \caption{The two maps used by Impetus to provide guidance and communicate AI results. (a) Attention map, where outlier patches and high uncertainty patches are highlighted in red, while other patches are in blue. The yellow recommendation boxes are generated by clustering attention values. (b) Prediction map, where red shows a high probability of tumor, and white shows a low probability of tumor, as predicted by the AI. The green and red boxes are areas of ``normal'' and ``tumor'', as labeled by the pathologist. Recommendation boxes generated by clustering attention values are also visible on this map.}
    \label{maps}
\end{figure}

In our communication with our pathologist collaborator, we learned that one major limitation of pathologists is the ability to efficiently and effectively navigate a large, high-resolution WSI. To address this limitation, we design AI to guide pathologists' attention to regions of major outliers that appear visually different from the rest of the WSI and are more likely to be tumors. Such guidance is manifested in two user interface elements:

\one {\bf Attention map} visualizes each patch's degree of outlying overlaid on the current WSI (Figure \ref{maps}(a)); \two {\bf Recommendation boxes} as a more explicit means to draw pathologists' attention to large clusters based on outlier detection results (Figure \ref{maps}(a), yellow box) --- these boxes are always visible, whether on the original WSI, on the attention, or on the prediction map (described below).

\begin{figure}
    \centering
    \includegraphics[width=1.0\linewidth]{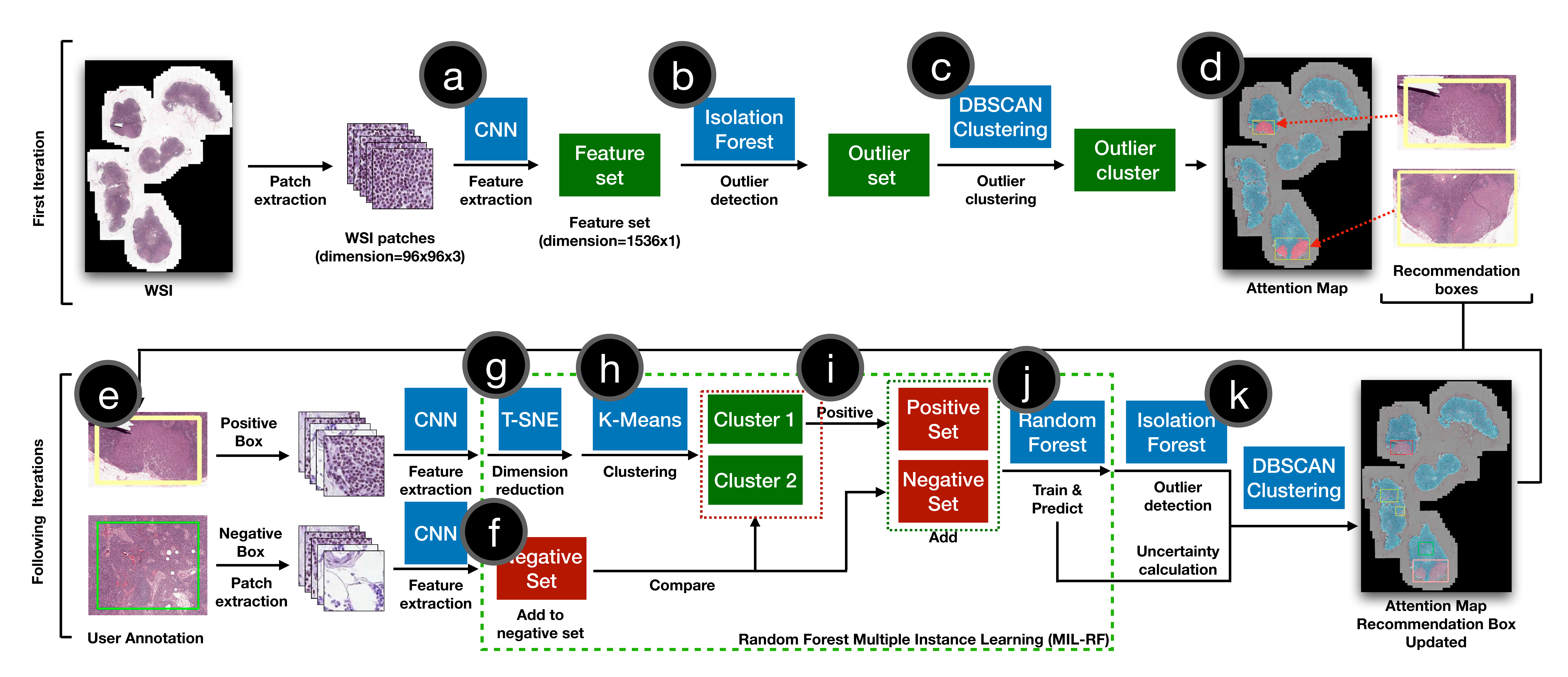}
    \caption{Overview of Impetus' AI backend. In the first iteration, Impetus first extracts the WSI to non-overlapping patches, followed by (a) feature extraction with a pre-trained CNN (InceptionResNetv2) model; (b) outlier detection by the isolation forest algorithm; (c) outlier clustering with the DBSCAN algorithm. Then, (d) an attention map with outlier clusters and recommendation boxes are generated. In the following iterations, the user first (e) annotates the recommendation boxes with agile labeling. Next, Impetus processes negative annotations by (f) adding negative box features to the negative set. For the positive annotation, Impetus uses (g) T-SNE to reduce the dimension of positive box features and applies (h) K-Means clustering to split them into two clusters. After that, Impetus (i) assigns the two clusters with labels by comparing them to the negative set and only adds features in the positive cluster to the positive set. Last, (j) a random forest classifier learns from the positive and negative set and predicts at a whole-slide level. The attention map and recommendation boxes are generated by (k) clustering from a combination of outliers and uncertain predictions. \rev{}{Procedures between (e - k) are repeated until the doctors are satisfied with the AI performance. }}
    \label{fig:flow_chart}
\end{figure}

\insec{Implementation} When the system is first loaded, a pre-trained InceptionResNetv2 model\footnote{We trained this model with image augmentation preprocessing, Adadelta optimizer with initial learning rate 0.1, binary cross entropy loss, 100 iterations with early stopping on validation loss.} \cite{ioffe2015batch} on PatchCamelyon dataset\footnote{\url{https://patchcamelyon.grand-challenge.org/}} extracts patch features (patch dimension=$96\times 96\times 3$, feature dimension=$1536\times1$) in WSIs (Figure \ref{fig:flow_chart}(a)). Given the imbalance nature of tumor {\it vs.} normal tissues, in the first iteration, the system performs isolation forest (max\_samples=256) \cite{liu2008isolation} outlier detection based on extracted features (Figure \ref{fig:flow_chart}(b)), and the detected outliers are highlighted in the attention map. 
In the following iterations, the attention map is a combination of outliers (from the initial detection) and high uncertainty patches (from specific models in each iteration)\footnote{Uncertainty is calculated as $\text{Uncertainty} = 1 - |0.5-\text{Probability}|\times 2$. The attention maps in the following iterations are calculated as the soft-OR of outliers and uncertainty: $\text{Attention} = \text{Uncertainty} \odot \text{Outlier}$.}. In order to obtain the recommendation boxes, the system uses a DBSCAN clustering algorithm (min\_sample=10, epsilon=3) \cite{ester1996density} to cluster WSI patches with attention value (Figure \ref{fig:flow_chart}(c,k)). In order to reduce users' distraction, the recommendation boxes are selected as the two clusters \rev{which}{that} occupy the largest areas on the WSI in each iteration.

\subsection{AI Using Agile Labeling to Train and Adapt Itself On-the-fly}
In digital pathology, the main challenge for AI is that, unlike other imaging modalities (\eg X-ray, CT), histological data (\eg ovarian carcinoma) tends to have a high variance across slides of different patients (sometimes same patients as well) \cite{kobel2010diagnosis}. Thus a pre-trained model often struggles to generalize to new data. To address this limitation, Impetus enables pathologists to use agile labeling to train AI on the fly.

{\bf Agile labeling} allows a pathologist to directly label on recommendation boxes (\fgref{walkthrough}(a)), or to draw a bounding box of tumor-negative patches (\fgref{walkthrough}(b)), or a box containing a mix of negative and positive cells, which serve as labels to train an existing model further to incorporate pathologists' domain knowledge. Importantly, such labeling technique is designed to be agilely achievable without incurring significant extra effort that interrupts the main diagnosis workflow.

\insec{Implementation} Agile labeling does not specifically require users to provide the exact contour of tumor tissues in a WSI, as strongly-supervised learning does. Alternatively, a user can marquee-select a positive box over an area which contains \textit{at least one} tumor patches, or a negative box on \textit{all negative} regions. We implemented a weakly-supervised MIL \cite{zhou2004multi, babenko2008multiple} to learn over such agile labels. 
To train the model, the system first initializes a $positive\_set$ and $negative\_set$. For each box annotated by a user, Impetus first partitions the WSI areas into $96\times 96 \times 3$ non-overlapping patches, and extracts the feature of each patch by the pre-trained CNN model from Section \ref{subc_outlier} ((Figure \ref{fig:flow_chart}(a))). Here, we denote each the extracted feature set as $X_i$ and the box-level annotation from user as $Y_i$.\footnote{In the MIL setup, each box only has one box-level annotation $Y_i$.}
For a negative box, all the patch features in the box can be included in $negative\_set$ (Figure \ref{fig:flow_chart}(f)). For positive boxes, the system uses T-SNE \cite{maaten2008visualizing} to represent the high-dimension features $X_i$ with two-dimension embedding $\overline{X}_i$\footnote{The embedding $\overline{X}_i$ is used for clustering for two reasons: \one avoiding K-Means to process high-dimension data, which could prevent clustering performance degradation; \two better visualizing the high-dimensional embedding space.} (Figure \ref{fig:flow_chart}(g)). Then, K-Means clustering is used to split $\overline{X}_i$ into two clusters: $\overline{X}_i^{(1)}$, $\overline{X}_i^{(2)}$ (Figure \ref{fig:flow_chart}(h)). After clustering, the algorithm compares the two clusters with negative samples from $negative\_set$  to pick the real positive cluster. After the positive cluster is recognized, all the instances in the positive cluster are included in the $positive\_set$ (Figure \ref{fig:flow_chart}(i)). Finally, a random forest classifier (MIL-RF, 100 trees, max\_depth=100)  \cite{breiman2001random} is trained with the obtained $positive\_set$ and $negative\_set$\footnote{The random forest algorithm is used since its ``resistance to overfit'' \cite{nalisnik2017interactive}. The notion of random forest has been applied to active learning \cite{nalisnik2017interactive}, or multiple instance learning algorithms \cite{leistner2010miforests}.} (Figure \ref{fig:flow_chart}(j)), and the user can continuously provide more annotations until the trained classifier reaches a satisfactory level of performance.

\subsection{AI Taking Initiatives Appropriately for the Level of Performance Confidence}

Even with agile labeling, lightweight on-the-fly learning only has limited improvement compared to training extensively offline. Thus it is crucial to convey the level of AI's performance to the pathologists. In Impetus, AI takes initiatives appropriately for its performance confidence level, as manifested in the following two user interface elements:
\one {\bf Prediction map} visualizes current AI's results overlaying the WSI, which serves to inform both the labeling and the usage of the current AI's model (Figure \ref{maps}(b)).
\two {\bf Initiatives based on confidence}---the more uncertain the AI `feels' about a WSI, the less initiative it takes, as shown in Table~\ref{tb:ai_initiatives}.

\insec{Implementation} Impetus has a rule-based confidence-level classifier to sort slides into three categories: high-confidence,  mid-confidence, and low-confidence. First, predictions of all the patches in the WSI are obtained. A patch has two characteristics: \textit{is\_positive} and \textit{is\_uncertain}. 
A patch is positive if the MIL-RF classifier output > 0.5, and is uncertain if MIL-RF classifier output $\in [0.25, 0.75]$. We empirically summarize the confidence-level decision rules\footnote{... which can be easily modified as a configuration of our tool.} as follows:

\begin{itemize} [leftmargin=0.25in]
    \item If there are more than 200 positive patches AND the number of positive patches is greater than twice the number of uncertain patches, then the slide is predicted as high-confidence;
    \item Else if there are no outlier clusters, then the slide is predicted as low-confidence;
    \item Else if the number of uncertain patches is greater than 300, then the slide is predicted as low-confidence;
    \item Else if the number of positive patches is greater than 200, then the slide is predicted as high-confidence;
    \item For all other cases, the slide is predicted as mid-confidence.
\end{itemize}

%% file: 05_study.tex
\section{Work Sessions with Pathologists}
To validate our design of Impetus, we observed how pathologists used this tool to perform diagnosis on a clinical dataset \cite{litjens20181399}. Our goal is to study whether the AI in Impetus \one can be compatibly integrated into pathologists' workflow and \two can provide added values to pathologists' diagnosis process.

\insec{Participants}
We recruited eight medical professionals from the pathology department in \rev{a local medical center}{UCLA Health}. The participants have experiences ranging from 1 to 43 years, including residents, fellows, and attending pathologists.

\insec{Data \& apparatus}
We used the Camelyon 17 \cite{litjens20181399} dataset and selected 16 WSIs\footnote{Our pilot studies indicated that 16 is the number of WSIs that would allow us to finish the session in about an hour to most effectively use the pathologists' time.} that were collected in the same medical center. Participants interacted with Impetus on a 15-inch laptop computer using a wired mouse. Impetus ran on a Microsoft Windows 10 Operating System using an Nvidia 960M GPU and 16GB RAM.

\insec{Design}
Our discussion with pathologists collaborators and an initial screening survey indicated that there was not a commonly-used digital pathology tool among the participants. To help pathologists calibrate their experience with Impetus, we introduced another tool --- ASAP\footnote{\url{https://computationalpathologygroup.github.io/ASAP/}}, which represents a very basic manual tool for viewing and annotating digital pathology slides. Each pathologist interacted with both Impetus and ASAP, which were referred to as System A and System B, respectively, to avoid biasing the pathologists. The order of tools was counterbalanced across the eight pathologists. Twelve of the 16 slides were diagnosed using Impetus and the remaining four using ASAP: we chose to keep more slides for Impetus as it was the target of our study, whereas ASAP was just to calibrate pathologists' tool experience.

\begin{figure}[h]
    \centering
    \includegraphics[width=0.6\linewidth]{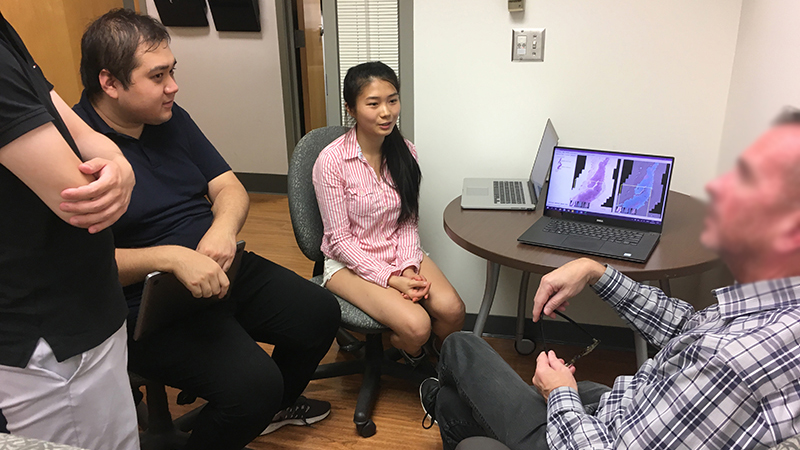}
    \caption{We conducted work sessions with eight pathologists from a local medical center to observe how they used Impetus as part of their diagnosis process.}
    \label{study}
\end{figure}

\insec{Tasks \& Procedure}
After briefly introducing the background of computer-assisted diagnosis, we walked each pathologist through a tool and let them practice on a separate toy dataset also gathered from \cite{litjens20181399}. We then asked questions about how the participant understood different interactive components, whether the tool was easy to learn and use, and whether the tool was helpful to their diagnosis. Then, the primary task began, which was to diagnose the entire group of WSIs using the provided tool in each condition. A trial started with a participant clicking to open a WSI and finished when they selected a diagnosis and clicked the `Confirm' button. After each condition, we further conducted a brief semi-structured interview for each participant to summarize their experience, feedback, and suggestions for the tool. Participants took a short break between the two conditions.

\insec{Analysis}
We employed an iterative open-coding method to analyze the qualitative data collected from the semi-structured interviews with pathologists. Two experimenters coded each participant’s data within one day after the study. One experimenter performed the first pass of coding and updated a shared codebook, which was then reviewed by the other experimenter to resolve disagreements. The two experimenters alternated the roles of the first coder and reviewer. After all the participants’ data were coded and consolidated, a third experimenter reviewed all the codes and transcripts and resolved disagreements through discussion with the previous two experimenters. Finally, we arrived at six high-level themes, which we summarize below as lessons learned.

%% file: 06_lessons.tex
\section{Findings, Lessons Learned \& Design Recommendations}

Based on the observations and data from the work sessions with pathologists, we present our findings below, which are summarized into six lessons. To maintain consistency, we organize these lessons using the same structure as {\S}5.

\subsection{AI Guiding Pathologists' Attention to Regions of Major Outliers}
\subsubsection{Recommendation boxes} (\fgref{walkthrough}a) were the most frequently used and discussed features during the study. We observed that in almost all the trials, pathologists started by zooming into the recommendation boxes and tried to provide annotations of the outlined region. Pathologists found it helpful to have such concrete start points in their examination.
 
\quo{... [recommendation boxes] narrow down the area of interest ... it helps}{7}

\quo{It was less effort because I was focusing only on the attention areas and not focusing on the other areas of the node so it was different from my usual way of looking at a slide.}{2}

However, pathologists did not always find the recommended regions matched their intuition, and they could not understand why certain regions were recommended.

\quo{... [the recommendation box] seems a little bit random. It's not necessarily areas that I would [look at] ...}{5}

\quo{The things it's focusing on does not correlate with at least what my brain thinks I am looking for.}{6}

A lack of transparency is not a new problem in recommender system research (\eg \cite{zhang2014explicit}). When introducing Impetus, we did explain that recommendation boxes were based on a detection of visual outliers, and all pathologists acknowledged that they understood such a concept. Although such outliers were computed based on histological features (the PatchCamelyon dataset), they did not always agree with what pathologists intuited as `interesting' regions worth examination.
When such a mismatch occurred---\ie an unexpected case of recommendation, pathologists could no longer reason about the recommendation boxes simply by referring to the abstract concept of `visual outliers'. At times, pathologists started to develop their own hypothesis of how AI was processing the WSI: \inquo{... it's interesting that it's picking area with fat as area of interest.}{2}

\lone
\vskip -0.5em
{\bf Recommendation \#1: an overview + instance-based explanation of AI's suggestions.}
Currently, Impetus only provides an explanation of the suggested regions at the overview level: a textual description of the outlier detection method as part of the tutorial and visualization (\ie attention map) that shows the degree of `outlying' across the WSI. As an addition, we can further incorporate instance-based explanation, \ie with information specific to a particular patient and a particular region on the patient's slide. The idea is to allow pathologists to question why a specific region is recommended by clicking on the corresponding part of the slide, which prompts the system to show a comparison between the recommended region and a number of samples from non-recommended parts of the slide for the physician to contrast features in these regions extracted by AI. One important consideration is that such an additional explanation should be made available on-demand rather than shown by default, which could defeat the recommendation boxes' purpose to accelerate the pathologists' examination process.
\vskip \baselineskip

We also found that pathologists wondered what they should do about the area outside of the recommendation boxes:

\quo{So I just look at the ones in the [recommendation] square?}{7}

\quo{Am I supposed to assume the rest of it is normal? I don't have to go searching for the rest of the slides for [tumor]?}{2}

Pathologists understood the implication {\it in} the recommendation boxes, \ie to prioritize certain regions of a WSI and to serve as a `shortcut' in lieu of scanning the entire WSI. However, it was unclear what was the implication {\it outside} of the recommendation boxes. This is especially true when pathologists could not find signs of tumor in the recommended regions: the system did not continue to guide them on how to proceed with the rest of the WSI.

\ltwo
\vskip -0.5em
{\bf Recommendation \#2: make AI-generated suggestions always available (and constantly evolving) throughout the process of a (manual) examination.} For example, in Impetus, a straightforward design idea is to show recommendation boxes one after another. We believe this is especially helpful when the pathologist might be drawn to a local, zoomed-in region and neglect looking at the rest of the WSI. The always available recommendation boxes can serve as global anchors that inform pathologists of what might need to be examined elsewhere beyond the current view. This is an example of a multi-shot diagnosis behavior where each shot is an attempt to find tumor cells in a selected region.

\subsubsection{Attention map} (Figure \ref{maps}a) visualizes outliers detected by the AI --- the same information based on which the recommendation boxes were drawn. It was designed to complement recommendation boxes with a backdrop of detailed guidance. We expected pathologists to use the attention map similarly as the recommendation boxes, \ie to direct their attention to look for more outlying regions for examination. However, pathologists did not find attention map useful:

\quo{The attention map shows the same thing as the recommended box. The box is enough to direct my attention.}{2}

\quo{I don't really see the point of the attention map ... These two maps are redundant.}{4}

The main difference was that recommendation boxes cost less effort to process, while \rev{}{the} attention map needed to be navigated (\ie panned and zoomed and interpreted (\ie mentally `decoding' the color scheme). 
Further, recommendation boxes provided actionable information (\ie to look into this box first), while \rev{}{the} attention map is action-neutral. Given that pathologists' overall goal is to eliminate the amount of area to study, they tended to prefer less extra effort and information with clearer actionability.

\lthree

{\bf Recommendation \#3: weigh the amount of extra efforts by co-designing a system with target medical users, as different physicians have different notions of time urgency.} 
Emergency room doctors often deal with urgent cases by making decisions in a matter of seconds, and internists often perform examinations in 15-20 minutes per patient; oncologists or implant specialists might decide on a case via multiple meetings that span days. There is a sense of timeliness in all these scenarios, but the amount of time that can be budgeted differs from case to case. To address such differences, we further recommend modeling each interactive task in a medical AI system (\ie how long it might take for the user to perform each task) and providing a mechanism that allows physicians to `filter out' interactive components that might take too much time (\eg the attention map in Impetus). Importantly, different levels of urgency should be modifiable (perhaps as a one-time setup) by physicians in different specialties.

\subsection{AI Using Agile Labeling to Train and Adapt Itself On-the-Fly}
{\bf Prediction map} (Figure \ref{maps}b) visualizes current AI's diagnosis of the WSI and was designed to help the pathologists assess the model's performance and decide where they could provide more labels.

However, pathologists used \rev{}{the} prediction map differently than we expected. Pathologists would often zoom into recommendation boxes on the WSI, study the region for a few seconds, then switch to the prediction map for a few seconds, and switch back to WSI. They tended to use the prediction map as a tool to help them see if there is something `interesting' in the current zoomed-in region. Sometimes pathologists used the prediction map for double-checking their developing diagnosis:

\quo{That was all negative, and I didn't get a strong heatmap signal, so it was confirmatory and somewhat helpful.}{6}

Interestingly, how pathologists used the prediction map seemed to complement the recommendation boxes: while recommendation boxes told pathologists which region is worth looking at (\ie might contain tumors), prediction map confirmed pathologists' assumption when they thought a region was of little `interest' (\ie no signs of tumor).

\lfour

{\bf Recommendation \#4: use visualization to filter out information, \ie leverage AI's results to reduce information load for the physicians.}
An example would be a spotlight effect that darkens parts of a WSI where AI detects little or no tumor cells. Based on our observation that pathologists used AI's results to confirm their examination \rev{on}{of} the original H\&E WSI, such an overt visualization can help them filter out subsets of the WSI patches. Meanwhile, pathologists can also reveal a darkened region if they want to examine further AI's findings (\eg when they disagree with AI, believing a darkened spot has signs of tumor).

The unexpected usage of \rev{}{the} prediction map affected agile labeling, as we discuss below.

{\bf Agile labeling} (\fgref{walkthrough}b) allows a pathologist to label on a recommendation box directly, or to marquee-select a region to coarsely annotate as normal or tumor. In the introduction phase, all pathologists reported having no problem understanding the idea of continuously labeling WSIs to improve the AI:

\quo{This is actually adding more work for me, but I would be willing to add labels knowing I would be improving the model}{4}

However, during the tasks, we noticed that almost all the labels were drawn only based on the recommendation boxes. Only one pathologist actively searched \rev{}{for} other regions to draw and provide more labels. It seemed that recommendation boxes served as a prompt, and pathologists were unmotivated to label other regions if unprompted.

We believe one fundamental reason is a lack of feedback to inform pathologists how important their labels were to the model retraining. Without such feedback, it might have been unclear to pathologists whether they needed to provide labels at all, or how much labeling would be enough.

\quo{Do I need to add labels?}{6}

\quo{Should I have provided more labels?}{5}

We assume that once pathologists see how a prediction map contained inaccurate results, they would be motivated to provide more labels to improve the prediction. However, our observations show that pathologists were more likely to make a diagnosis directly by manual examination, instead of correcting AI's predictions as we expected. Falling back to manual examination seems a more cost-effective alternative to AI automation than \rev{trying to iteratively improve the AI}{ than improving the AI iteratively}.

\lfive

{\bf Recommendation \#5: when adapting the model on-the-fly, show a visualization that indicates the model's performance changes as the physician labels more data.}
There could be various designs of such information, from showing low-level technical details (\eg the model's specificity vs. sensitivity), high-level visualization (\eg charts that plot accuracy over WSIs read) and even actionable items (\eg `nudging' the user to label certain classes of data to balance the training set).
There are two main factors to consider when evaluating a given design: 
\one as we observed in our study, whether the design could inform the physician of the model's performance improvement or degradation as they label more data, which can be measured quantitatively as the amount of performance gain divided by the amount of labeling work done;
\two as we noted in Lesson \#2, whether consuming the extra information incurs too much effort and slows down the agile labeling process, and whether there is actionability given the extra information about model performance changes.

\subsection{AI Taking Initiatives Appropriately for the Level of Performance Confidence}
As shown in Table~\ref{tb:ai_initiatives}, AI's level of initiative is mediated based on its level of confidence about the model's performance.
For low-confidence cases, AI took no initiative, and all pathologists were mostly unaware of AI's presence, while they simply focused on performing the usual manual diagnosis.
For high-confidence cases, as expected, pathologists quickly confirmed AI's proactive diagnosis of macro --- the easiest type of tumor to detect by both pathologists and AI. However, when it comes to cases diagnosed as negative by the AI, pathologists tended to perform a manual diagnosis anyway:

\quo{On the ones that it said it's confident but didn't really tell you it's negative, I still felt like I had to look at those to confirm. I wasn't going to trust the system [to confirm] that it's negative}{2}

In pathology, in order to rule out tumors, pathologists have to thoroughly examine the entire WSI, whereas it only takes one positive case to diagnose the lymph node as positive. Thus there was a discrepancy of trust between macro vs. negative, despite that AI treats both equally as different labels of a slide image and categorizes both as high confidence.

\lsix

{\bf Recommendation \#6: provide additional justification for a negative diagnosis of a high-staked disease.}
For example, when Impetus concludes a case as negative, the system can still display the top five regions wherein AI finds the most likely signs of tumor (albeit below a threshold of positivity). In this way, even if the result turned out to be a false negative, the physicians would be guided to examine regions where the actual tumor cells are likely to appear.
Beyond such intrinsic details, it is also possible to retrieve extrinsic information, \eg prevalence of the disease given the patient's population, or similar histological images for comparison. As suggested in \cite{xie2020chexplain}, such extrinsic justification can complement the explanation of a model's intrinsic process, thus allowing physicians to understand AI's decision more comprehensively.

For the mid-confidence case, AI was designed to pre-fill the diagnosis dialog (but without any confirmative action) as a way to hint its prediction without signaling any conclusive decision. This design did not seem to have noticeable effects on the pathologists, which echos Lesson \#3 that information needs to present actionability in order to affect a medical user's workflow.

%% file: 07_discussion.tex
\section{Limitations, Discussion and Conclusion}

This paper explores how AI's capabilities with limitations can still benefit a traditional manual diagnosis process on histological data. We investigate this question through the design and study of Impetus, a tool where an AI takes multi-leveled initiatives to provide various forms of assistance to a pathologist performing tumor detection in whole-slide histological images. We conducted work sessions with eight medical professionals using Impetus and summarize our findings and lessons learned, based on which we provide design recommendations with concrete examples to inform future work on human-centered medical AI systems.

Due to the limited availability of pathologists, the number of participants (N=8) of this work was small, and all participants were from the same medical center. The performance of the multiple instance learning model was not evaluated because of the limited number of new labels obtained from pathologists in the work sessions, which we discussed in Lesson \#5 on engaging pathologists to provide extra input.

Below, we discuss several other limitations encountered during the development of our system.

{\bf Detecting small lesion tissues in a WSI}.
We found it hard for the system to detect small lesion tissues in a WSI during the work session. However, from our interviews with medical professionals, it is more valuable to find those areas with AI, whereas large, macro tissues can be located quickly without assistance. Here we summarize the reasons why the machine learning algorithm fails to localize small lesion tissues. First, the system takes patches as input to extract the features and classifies them with trained MIL-RF. However, this approach can be problematic when detecting small-area lesion tissues, since the classification performance highly depends on color, and small metastasis do not change the color of patches significantly. Further, the machine learning model treats tissues in a WSI as separate patches without considering structural correlations among the tissues. Specifically, lymph node invasion starts from the perimeters of the node. Thus small lesion tissues are more likely to appear in those peripheral regions. 

{\bf Integration into pathologists' workflow}.
Data encountered in a pathologist's day-to-day workflow are imbalanced by nature: metastasis areas often occupy small fractions of the entire WSI. As a result, there would be more negative annotations than positive ones. Furthermore, with the coarse labeling enabled by our MIL algorithm, only a subset of the patches in a positive annotation are truly positive patches. This imbalance in training data skews the model's predictions.

On the other hand, to use an AI system for diagnosis, the AI's performance must be validated. However, a trained-on-the-fly AI can not be practically validated, \ie it is not feasible to fully validate the model's performance after every iteration of labeling and training, even though the number of new training data is small. Such a dynamic system is hard to control to maintain a certain level of performance regardless of run-time user interaction.

We did not observe any automation bias (\ie physicians biasing their decision in favor of AI's output), primarily because \one the current AI's performance is imperfect and pathologists often did not trust the model to automate a diagnosis; \two we designed the AI to reframe from automation in cases of a low confidence, thus prompting pathologists to take control and rely on their own expertise.

{\bf Combining prior knowledge}.
In the real diagnosis environment of a pathologist, extra patient information is necessary for diagnosing a glass slide, which often includes the patient's medical history and type of cancer as determined from previous examinations. From our interviews with pathologists, we learn that such information is crucial for diagnosis speed and accuracy, as it informs the pathologist on what to look for and where to find them. To better match the AI with a pathologist's mental model and provide better guidance and explanations, we should incorporate patient information into the diagnosis model. For example, the AI might use a different CNN to look for a specific type of tumor tissues given a particular cancer type. 

{\bf Explicit vs. implicit feedback}.
So far, Impetus primarily relies on explicit feedback from physicians via agile labeling. Future work should leverage other information as implicit feedback, \eg what kinds of WSI areas a pathologist looks at first, or spends the most time examining, where and how much they zoom in. Inferring useful information for adapting the model presents new technical challenges for the machine learning community; for HCI and CSCW researchers, the challenge is making such inference transparent by informing pathologists how AI is learning from some of their implicit behavior.

%% file: 09_acknowledgement.tex
\begin{acks}
\rev{}{
We thank the reviewers for their valuable feedback. We thank all the anonymous participants for their participation in our study.}
This work was funded in part by a Hellman Fellowship and the National Science Foundation under grant IIS-1850183.

\end{acks}

%% file: 08_supplementary.tex
\newpage

\section{Implementation Details of Impetus Software}
We introduce the implementation of Impetus. Overall, the software was written in Python and the detailed implementation of Impetus' capabilities are shown as follows.

{\bf Attention map \& recommendation boxes} 
When the system is first loaded, Otsu method \cite{otsu1979threshold} is first utilized to separate foreground tissues and background. Then a pre-trained InceptionResNetv2 model \cite{ioffe2015batch} with PatchCamelyon dataset\footnote{\url{https://patchcamelyon.grand-challenge.org/}} extracts patch features in WSIs. We trained this model with image augmentation preprocessing, Adadelta optimizer with initial learning rate 0.1, binary cross entropy loss, 100 iterations with early stopping on validation loss. The input dimension of the pre-trained network is $96\times 96\times 3$, and the extracted features have a dimension of $1536\times 1$. In the first iteration, the system performs isolation forest \cite{liu2008isolation} outlier detection (max\_samples=256) based on extracted features, and the detected outliers are highlighted in the attention map. 
In the following iterations, the attention map is a combination of outliers (from the initial detection) and high uncertainty patches (from specific models in each iteration). Uncertainty is calculated as $Uncertainty = 1 - |0.5-Probability|\times 2$. The attention map is calculated as the soft-OR of outliers and uncertainty: $Attention = Uncertainty \odot Outlier$. In order to obtain the recommendation boxes, the system uses a DBSCAN clustering algorithm \cite{ester1996density} (min\_sample=10, epsilon=3) on the attention map to find clusters. In each iteration, Impetus only selects the two clusters which occupy the largest areas on the WSI as recommendation boxes to reduce users' distraction.

\begin{algorithm}[h]
\caption{Impetus MIL Training}
\label{alg:A}
\begin{algorithmic}
\STATE {initialize $positive\_set = [~]$, $negative\_set = [~]$}
\WHILE{Performance not satisfied}
\STATE{Annotate $X_1, X_2, \dots, X_N$ boxes with $Y_1, Y_2, \dots, Y_N$ labels}
\FOR{$X_i$}
\IF{$Y_i == -1$}
\STATE{$negative\_set.\text{append}(X_i)$}
\ELSE
\STATE{Embed $X_i$ to $\overline{X}_i$ with T-SNE}
\STATE{Split $\overline{X}_i$ into $\overline{X}_i^{(1)}$, $\overline{X}_i^{(2)}$ with K-Means, map the split to original data $X_i^{(1)}$,  $X_i^{(2)}$}
\STATE{Assign each instance in $X_i^{(1)}$ with $+1$ labels, and $X_i^{(2)}$ with $-1$ labels}
\STATE{Train a random forest classifier with $X_i^{(1)}$ and $X_i^{(2)}$}
\STATE{Predict negative box $X_{neg}$}
\STATE{Adjust the labels of $X_i^{(1)}$ and $X_i^{(2)}$}
\STATE{Append positive set with positive instances}
\ENDIF
\ENDFOR
\STATE{Train a random forest classifier with $positive\_set$ and $negative\_set$}
\ENDWHILE
\end{algorithmic}
\end{algorithm}

{\bf Agile labeling} Impetus uses an agile labelling technique that does not specifically require users to provide the exact contour of tumor tissues in a WSI, as strongly-supervised learning does. Alternatively, a user can mark a positive box over an area which contains \textit{at least one} tumor patches, or a negative box on \textit{all negative} regions. We implemented a weakly-supervised multiple instance learning (MIL) to enable a traditional random forest algorithm learn over such agile labels. As shown in Algorithm \ref{alg:A}, $X_i$ is a feature set in within a box and only has one box-level label $Y_i$. For a negative box, all the instances in the box can be included in $negative\_set$. In order to avoid the performance degradation of the following clustering algorithm, we first use manifold learning -- T-SNE \cite{maaten2008visualizing} -- to represent the 1536 dimension feature $X_i$ with two-dimension embedding $\overline{X}_i$. Here, it is assumed that the T-SNE would embed the positive box instances into one positive and one negative cluster, thus we use K-Means clustering to split $\overline{X}_i$ into two clusters: $\overline{X}_i^{(1)}$, $\overline{X}_i^{(2)}$. By using the split, the original 1536-\rev{dimention}{dimension} $X_i$ can be partitioned into $X_i^{(1)}$ and $X_i^{(2)}$. After clustering, the algorithm compares the two clusters with negative samples from negative box $X_{neg}$ to pick the real positive cluster. To achieve this goal, it first assigns each instance in $X_i^{(1)}$ with $+1$ labels, and that in $X_i^{(2)}$ with $-1$ labels, then trains a random forest classifier (100 trees, max\_depth=100) with $X_i^{(1)}$ and $X_i^{(2)}$. The trained classifier is used to predict negative box instances $X_{neg}$ previously provided by the user. Finally, the cluster which has the \textit{opposite} prediction to the negative box is the positive cluster. After the positive cluster is recognized, all the instances in the positive cluster are included in the $positive\_set$, and the instances in the rest cluster are aborted. Finally, a random forest classifier (MIL-RF) is trained with the obtained $positive\_set$ and $negative\_set$. Here, the random forest algorithm is selected since its ``resistance to overfit'' \cite{nalisnik2017interactive}. What's more, the notion of random forest has been applied to active learning \cite{nalisnik2017interactive}, or multiple instance learning algorithms \cite{leistner2010miforests}. In the following iterations, the user can iteratively provide more annotations until the trained classifier reaches a satisfactory level of performance.


{\bf Confidence calculation} Impetus has a rule-based confidence-level classifier to classify slides into three categories, namely high-confidence, mid-confidence and low-confidence.
The system first obtains the predictions of all the patches in the WSI. A patch has two characteristics: \textit{is\_positive} and \textit{is\_uncertain}. 
A patch is positive if the MIL-RF classifier output > 0.5, and is uncertain if MIL-RF classifier output $\in [0.25, 0.75]$. We empirically summarize the confidence-level decision rules as follows:

\begin{itemize}
    \item if there are more than 200 positive patches AND the number of positive patches is greater than twice the number of uncertain patches, then the slide is predicted as high-confidence;
    \item else if there are no outlier clusters, then the slide is predicted as low-confidence;
    \item else if the number of uncertain patches is greater than 300, then the slide is predicted as low-confidence;
    \item else if the number of positive patches is greater than 200, then the slide is predicted as high-confidence;
    \item for all other cases, the slide is predicted as mid-confidence.
\end{itemize}
